\documentclass[11pt]{article}
\usepackage{moriond,epsfig}
\usepackage{xspace}
\usepackage{url}

\def\Journal#1#2#3#4{{#1} {\bf #2}, #3 (#4)}

\def\PRD{{\em Phys. Rev.} D}

\def\be{\begin{equation}}
\def\ee{\end{equation}}
\def\bea{\begin{eqnarray}}
\def\eea{\end{eqnarray}}

\newcommand{\dzero}{D\O\xspace}
\newcommand{\mtop}{\ensuremath{m_{\rm top}}\xspace}
\newcommand{\mW}{\ensuremath{m_{W}}\xspace}
\newcommand{\mhiggs}{\ensuremath{m_{\rm Higgs}}\xspace}
\newcommand{\ljets}{\ensuremath{\ell}+\rm jets\xspace}
\newcommand{\ttbar}{\ensuremath{t\bar{t}}\xspace}
\newcommand{\GeVcc}{\ensuremath{\mathrm{Ge\kern-0.1em V/c^2}}\xspace}
\begin{document}
\vspace*{4cm}
\title{Top Quark Mass Measurements at the Tevatron and the Standard Model Fits}

\author{ M.H.L.S. Wang for the CDF and \dzero collaborations}

\address{Fermi National Accelerator Laboratory,\\
Batavia, IL 60510, U.S.A.}

\maketitle\abstracts{
New measurements of the top quark mass from the Tevatron are presented. 
Combined with previous results, they yield a preliminary new world average of
$\mtop=170.9\pm1.1({\rm stat})\pm1.5({\rm syst})\GeVcc$ and impose new
constraints on the mass of the Higgs boson.}

\section{Introduction}

The huge interest in a precise measurement of the top quark mass (\mtop) is
primarily motivated by its role in constraining the mass of the Higgs boson
(\mhiggs).  To see this, let us begin by looking at the
mass of the $W$ boson (\mW) in the Standard Model which, when one-loop
radiative corrections are included, can be related to well known electroweak
quantities through the following expression:
\bea
  m_{W}^{2} = \frac{\frac{\pi\alpha}{\sqrt{2}G_F}}{\sin^2\theta_W\left(1+\Delta
  r\right)}.
\eea
The radiative corrections contained in $\Delta r$ receive contributions from the top
quark:
\bea
  {(\Delta r)}_{\rm top} \approx \frac{3G_F m_{\rm top}^{2}}{8\sqrt{2}\pi^2}
  \frac{1}{\tan^2\theta_W}
\eea
and the Higgs boson:
\bea
  {(\Delta r)}_{\rm Higgs} \approx 
  \frac{11G_F m_{Z}^{2}\cos^2\theta_W}{24\sqrt{2}\pi^2}
  \ln{\frac{m_{\rm Higgs}^{2}}{m_{Z}^{2}}}
\eea
where $m_{Z}$ is the mass of the $Z$ boson. From these expressions, we see that
\mtop enters quadratically while \mhiggs enters logarithmically.  A precise
knowledge of both \mW and \mtop in combination with existing electroweak data
is therefore necessary to impose useful constraints on \mhiggs.  Such
constraints, in turn, are of tremendous value in the ongoing search for the
Higgs.

In this talk, we present the latest top quark mass measurements from the CDF
and \dzero collaborations based on up to 1 fb$^{-1}$ of Run II data collected
at Fermilab's Tevatron. These results are combined with previous ones to give a
new preliminary world average for \mtop which, in turn, yields new constraints
on the Higgs mass.

\section{Measurement Channels and Experimental Challenge}

Now that we understand the motivation behind a precise determination of the top
mass, let us look at the top quark decay channels in which these measurements
are performed and the experimental challenges they pose.

In the all jets channel, both $W$ bosons from the \ttbar pair decay
hadronically into jets for a total of 6 jets in the event.  This channel has
the advantage of having the largest branching ratio of 44\%.  It suffers, however, from large background levels from QCD multijet
events.  On the other hand, it benefits from the presence of the hadronically
decaying $W$ bosons whose well known masses can be exploited to perform an
in-situ calibration of the jet energies, reducing the effect of the
systematic uncertainty in the overall jet energy scale.
In the dilepton channel, both $W$ bosons decay leptonically.  It has the
advantage of having the lowest background levels coming from Drell-Yan
processes associated with jets, diboson production with associated jets, and
$W$+3 jet events with one jet faking an electron.  Unfortunately, it also has
the lowest branching ratio of 5\%.
In the lepton+jets (\ljets) channel, one of the two $W$ bosons from the \ttbar
pair decays hadronically while the other one decays leptonically.  This channel
maintains a good balance between a reasonable branching ratio of 29\% and
moderate background levels from $W$+jets and QCD multijet events.  Like the all
jets channel, it can benefit from an in-situ jet energy calibration using the
\mW constraint.  It has traditionally yielded the most precise \mtop
measurements.

To appreciate the challenge involved in measuring the top mass at the Tevatron,
let us now take the \ljets channel as an example.  In this case, what our
reconstruction programs give us from the detector are several jets, a high
$p_T$ lepton, substantial missing transverse energy, and an interaction vertex.
Since we don't really know how to associate jets with partons in general, all
jet permutations need to be considered in a straightforward reconstruction of
the top mass.  Furthermore, unlike long lived particles, there are no detached
vertices associated with the top quark itself that can be used to separate the
signal from the background events.  This means that, even with $b$-tagging, 
there are no sharp and clean mass peaks from which the top mass can be
determined directly.  Fortunately, despite these challenges, sophisticated
measurement techniques have been developed that make a precise measurement of
the top mass possible.

\section{Top Quark Mass Measurement Techniques}

In this section we describe the three major techniques used in measuring
the top quark mass.  All the measurements presented here use one or some
combination of these techniques.

The template method is the oldest of the three techniques and has been used for
most of the earliest mass measurements.  In this technique, one begins by
identifying a variable sensitive to the top mass, an obvious choice of which
would be the kinematically reconstructed value of the mass itself. 
Distributions of the chosen variable are then plotted separately for several
samples of fully simulated Monte Carlo (MC) events differing only in the value
of the top mass used to generate the signal events.  Each of these distributions is
called a template and is associated with a particular value of the input mass.
The top mass is then extracted from the data sample by comparing the data distribution
directly with each MC template to find the best fit value based on some measure
of the goodness of fit.  More recent applications of this technique
parameterize the templates in terms of a probability density function which is
used to construct likelihoods from which the top mass is extracted.

\dzero pioneered the  application of the matrix element (ME) method to top
quark mass measurements in the Run I data from the \ljets channel
\cite{nature}.  It is based on calculating the probability for observing each
event which includes contributions from both signal and background sources. The
signal probability is calculated as a function of the assumed top mass,
resulting in a probability distribution for each event.  The probability is
taken to be the differential cross section for the process in question.  The
calculated probability distributions for every event in the data sample are
combined to construct a joint likelihood from which the top mass is determined
and its uncertainty estimated.  The ME method makes use of as many measured
variables as possible to completely specify an event, thereby allowing maximum
discrimination between signal and background events.  Within each event, all
possible jet permutations are combined in a natural way based on their relative
probabilities.  Furthermore, the use of transfer functions allows a
probabilistic treatment of the mapping between parton and jet energies where
the full spectrum of parton energies contributing to the observed jet energy is
taken into account.

The ideogram method, like the ME method, calculates an event-by-event
likelihood.  This technique makes use of a constrained kinematic fit to
reconstruct the top mass.  Using a simple parameterization, the  probability for
observing the reconstructed mass is then calculated as a function of the true
value with the measurement resolution taken into account.  This technique,
which was also pioneered by \dzero \cite{dzeroideogram}, aims to achieve
statistical uncertainties comparable to those of the ME method without
requiring as many computational resources.

\section{New Results from the Tevatron}

\begin{figure}
\begin{center}
\includegraphics[height=2.2in]{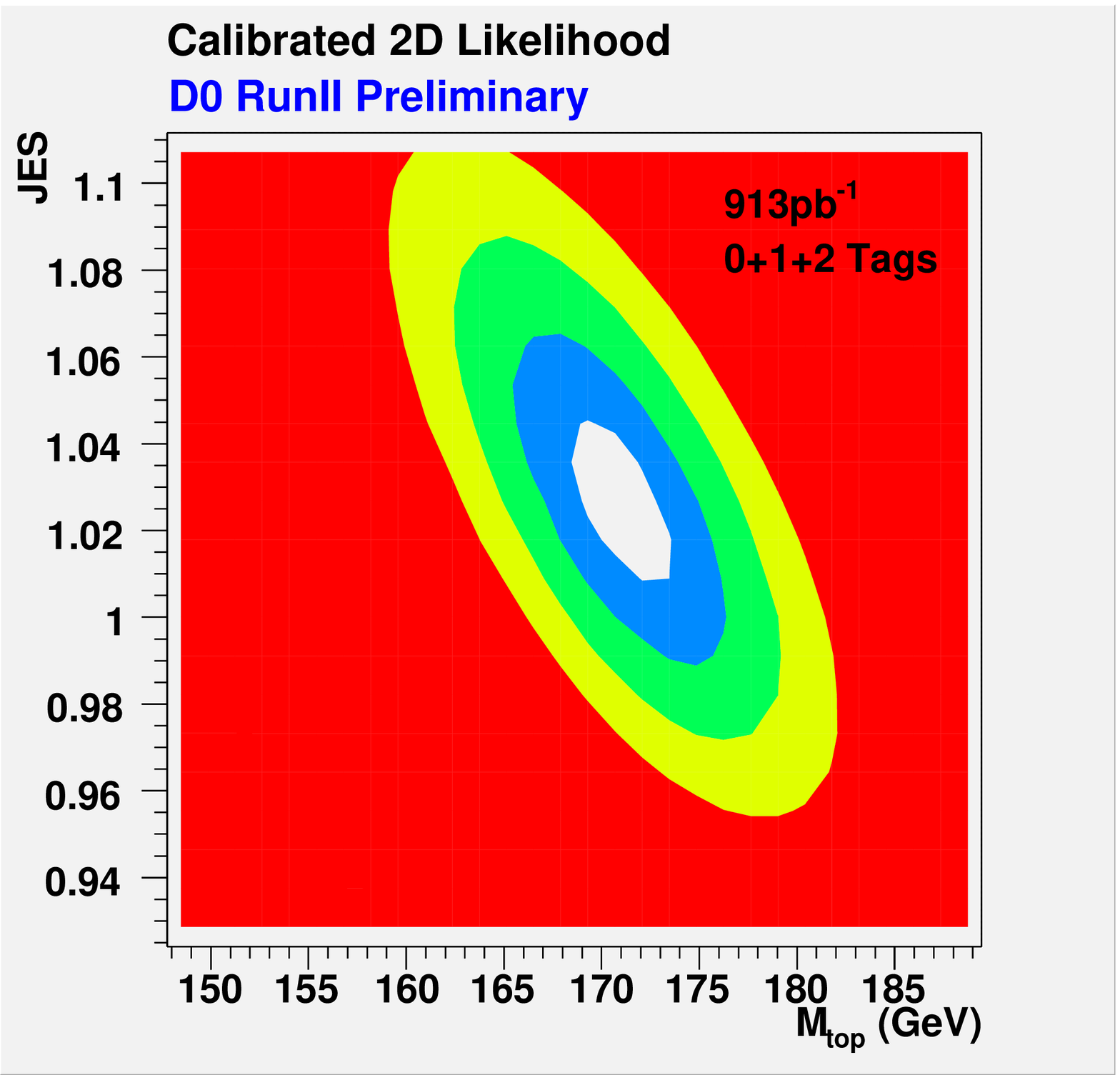}
\includegraphics[height=2.2in]{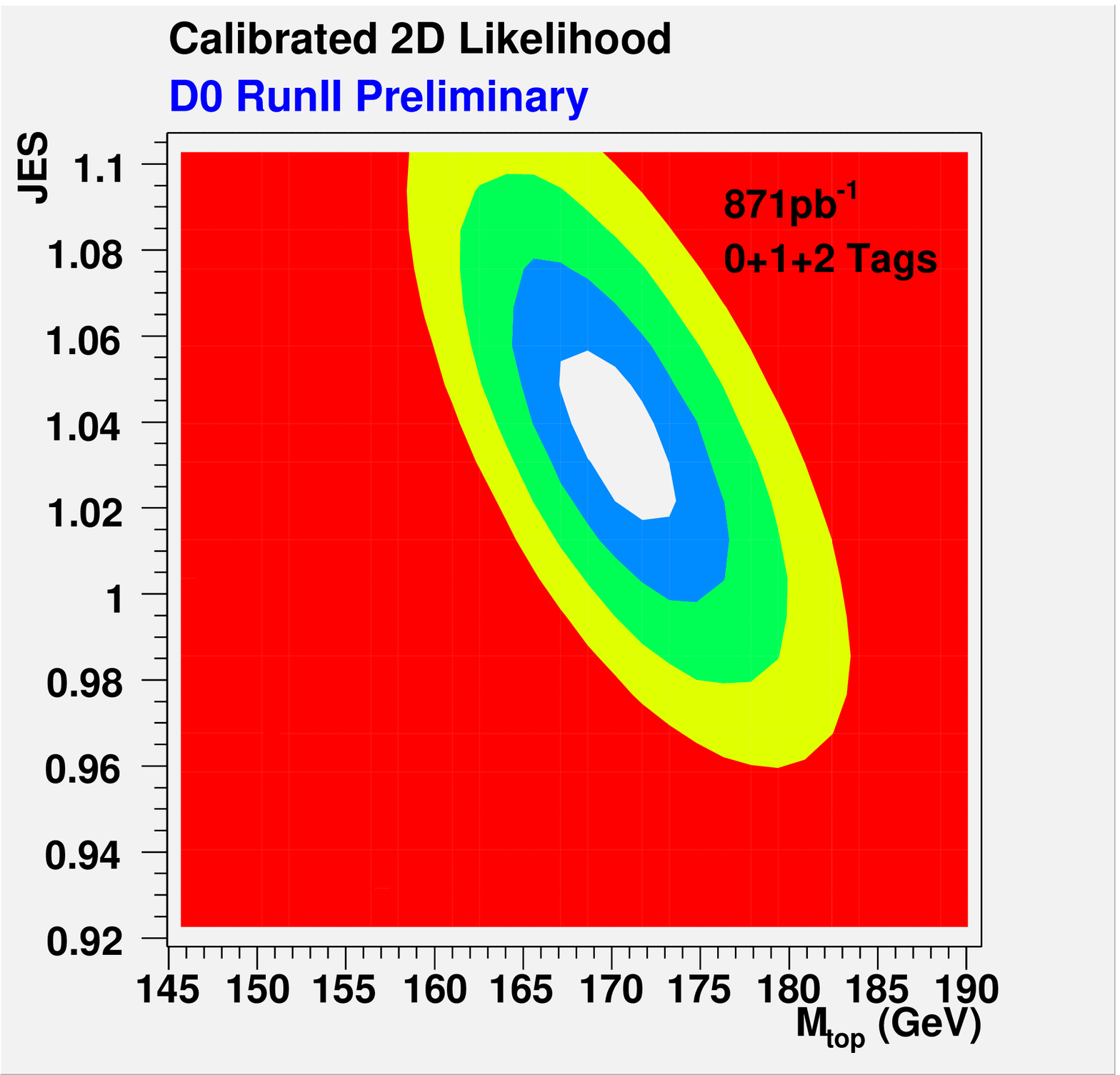}
\caption{2D likelihoods for electron and muon channels for the \dzero \ljets
result.
\label{fig:dzero-ljets}}
\end{center}
\end{figure}
\dzero has measured the top quark mass in the \ljets channel using the ME method
described in the previous section \cite{cdfljets}.  This measurement takes
advantage of the \mW constraint to perform an in-situ calibration of the jet
energies.  This is done by introducing a global scale factor, $JES$, that is
applied to the energies of all the jets.   A fit is then performed that
maximizes the likelihood simultaneously in \mtop, $JES$, and the signal
fraction $C_s$.  The 2D likelihood fits in \mtop and $JES$ are shown separately
for the electron and muon channels in Figure \ref{fig:dzero-ljets}.  The
combined result for both channels is $170.5\pm2.4({\rm stat+JES})\pm1.2({\rm
syst})\GeVcc$ for 0.9 fb$^{-1}$ of data.  Dominant systematic uncertainties are
in the modeling of initial and final state radiations and $b$-fragmentation.
This is the best \dzero measurement of the top quark mass to date.

\begin{figure}
\begin{center}
\includegraphics[height=2.2in]{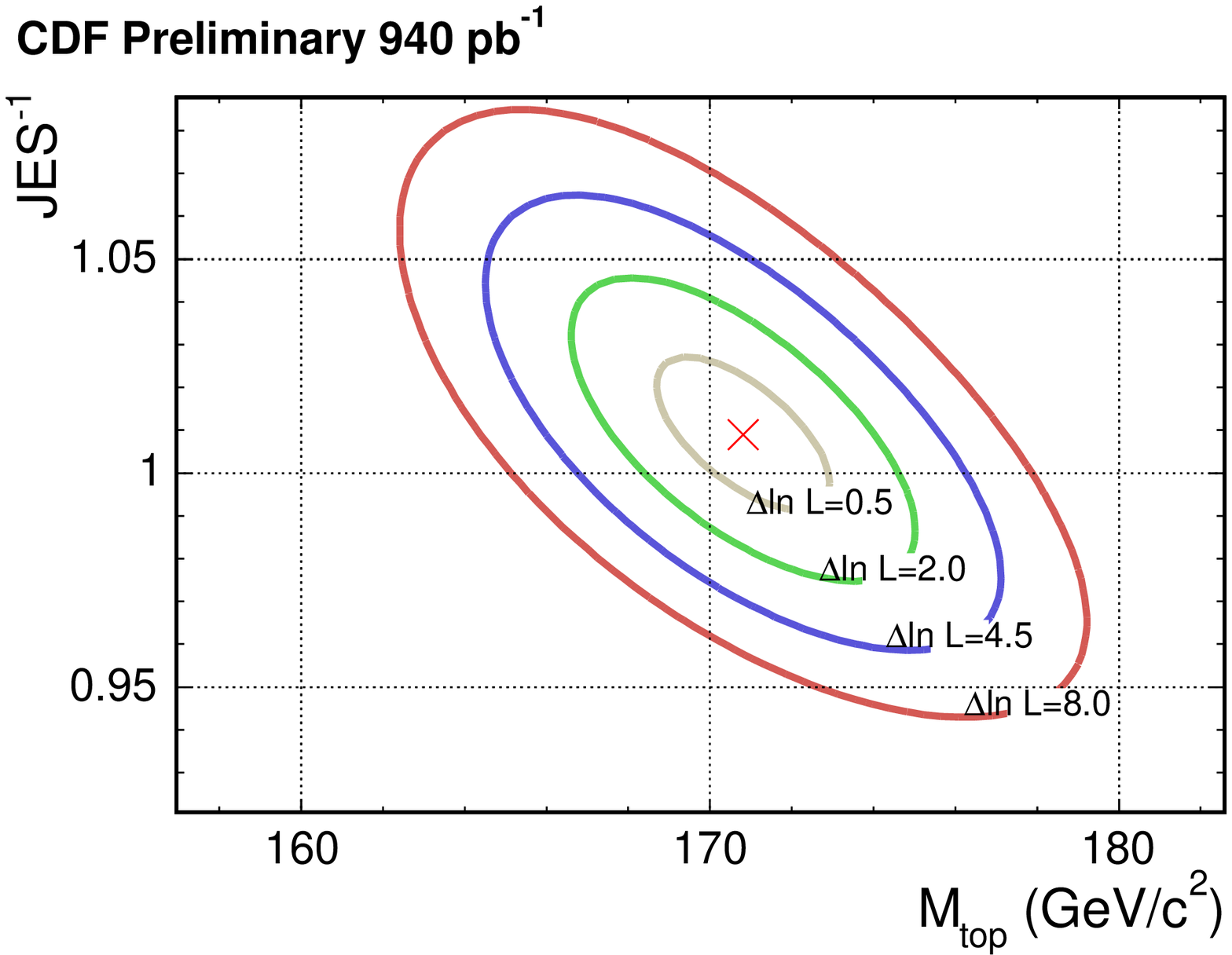}
\includegraphics[height=2.2in]{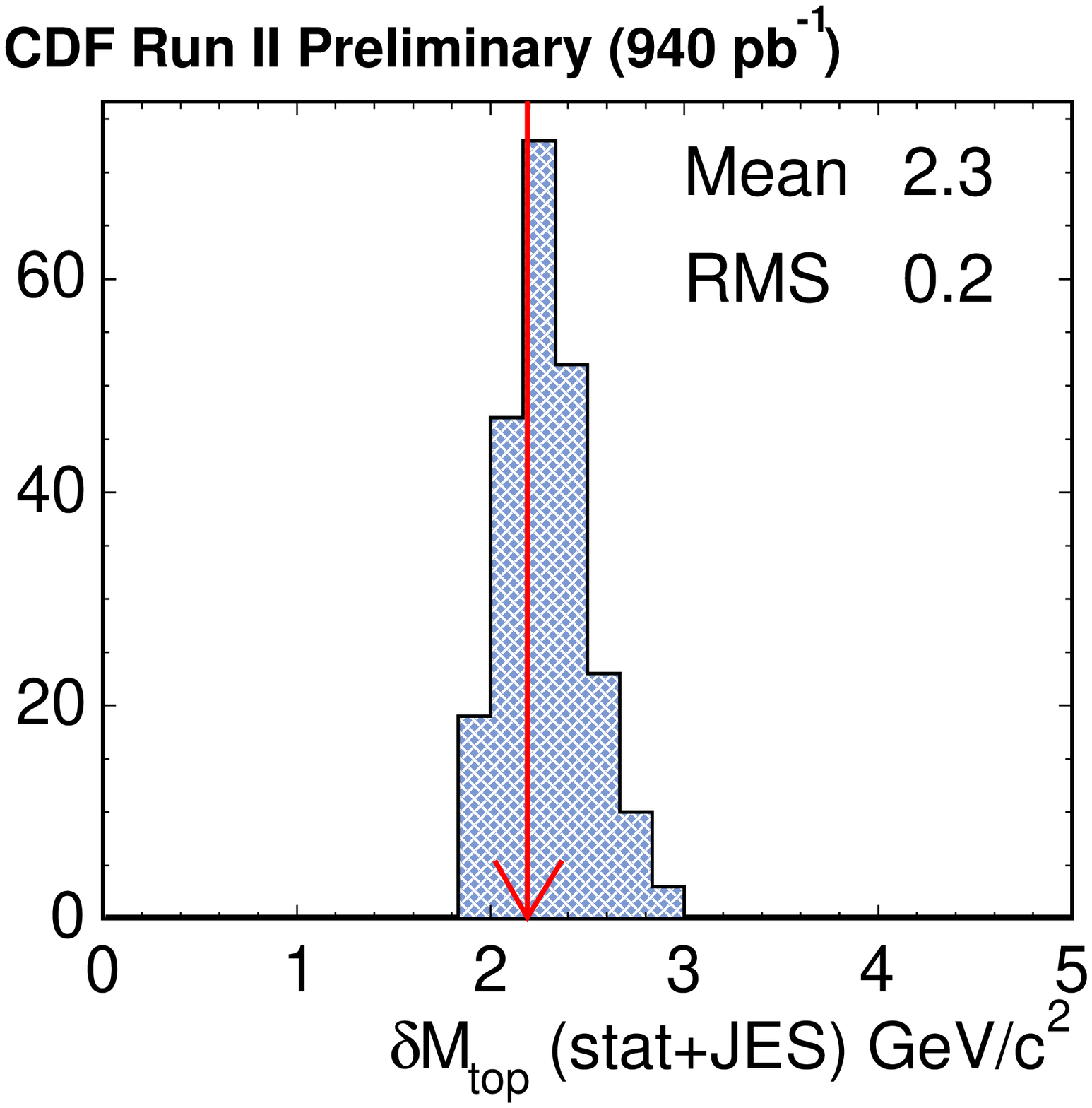}
\caption{Likelihood fit to data and expected error distributions for the CDF \ljets
result.
\label{fig:cdf-ljets}}
\end{center}
\end{figure}

CDF has also measured the top quark mass in the \ljets channel using the ME method.  Like
the \dzero result, this measurement employs an in-situ jet energy  calibration
through the inclusion of a global $JES$ parameter in the likelihood fit.  The
left plot in Figure \ref{fig:cdf-ljets} shows the 2D likelihood fit to the data
for both electron and muon channels in $JES$ and \mtop.  The right plot in
Figure \ref{fig:cdf-ljets} shows the expected error distribution from MC
ensemble tests with the arrow indicating the measurement uncertainty.  The
measured result for 0.94 fb$^{-1}$ of data is $170.9\pm2.2({\rm
stat+JES})\pm1.4({\rm syst})\GeVcc$.  The largest systematic uncertainty is in
the modeling of initial and final state radiations. This is currently the
most precise CDF measurement of the top quark mass. 

\begin{figure}
\begin{center}
\includegraphics[height=2.2in]{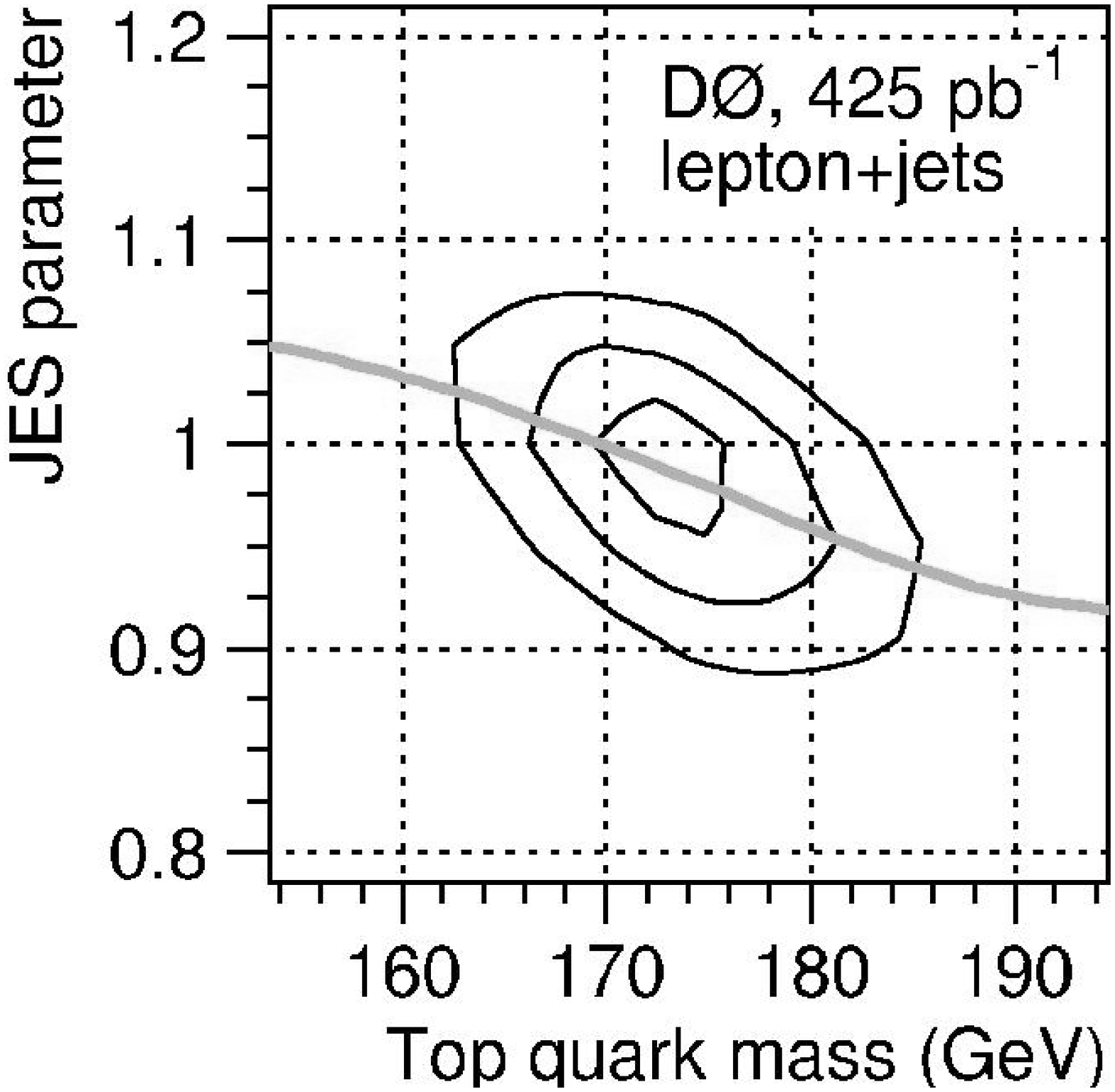}
\includegraphics[height=2.2in]{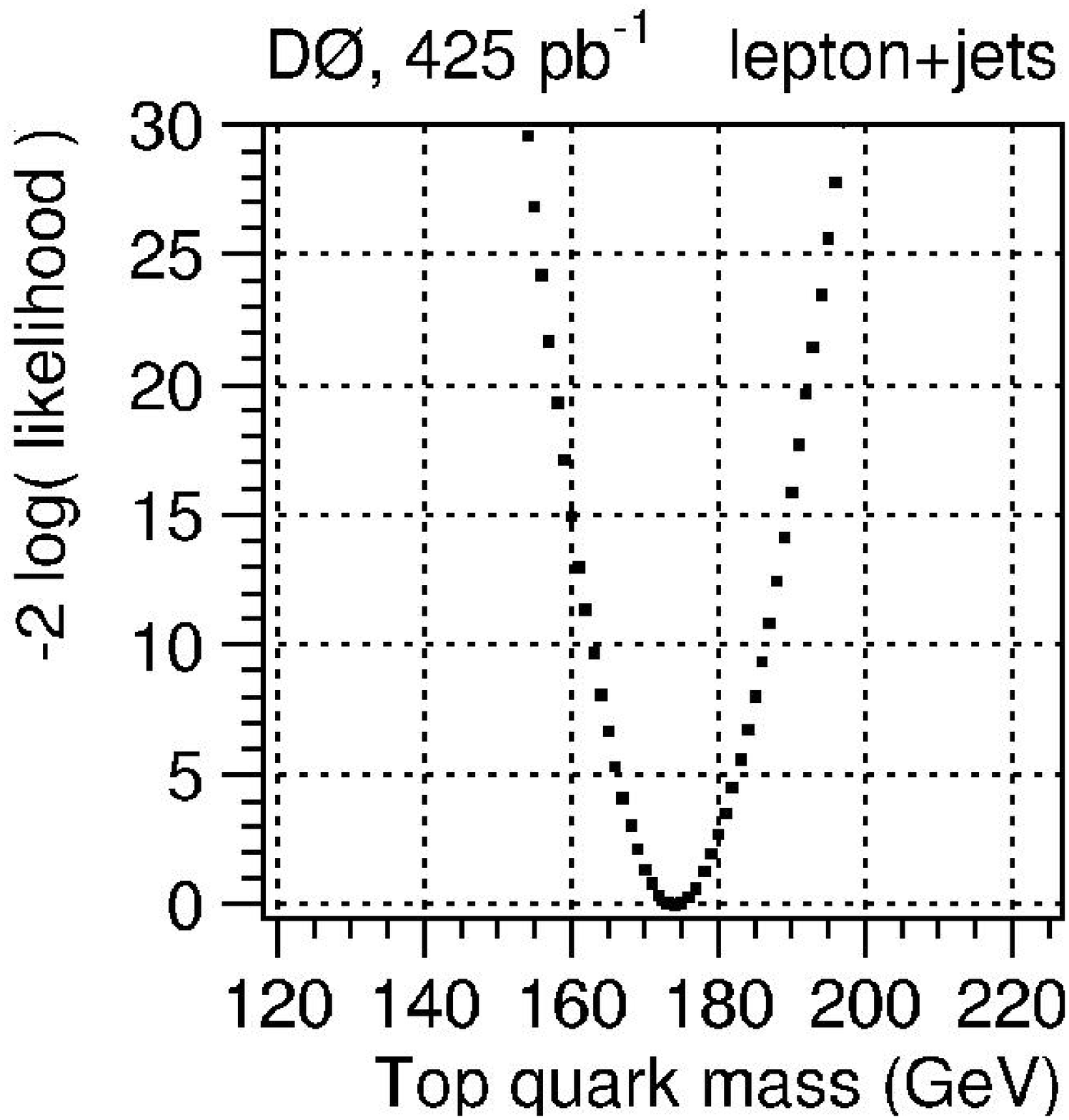}
\caption{2D and 1D likelihoods for the \dzero ideogram \ljets results.
\label{fig:dzero-ideogram}}
\end{center}
\end{figure}

\dzero has a measurement of the top quark mass in the \ljets channel using the ideogram
method \cite{dzeroideogram}.  Like the two results above, it employs an in-situ
jet energy calibration.  The 2D likelihood as a function of $JES$ and \mtop is
shown on the left in Figure \ref{fig:dzero-ideogram} with the gray line
indicating the fitted value of $JES$ as a function of \mtop.  The right plot in
Figure \ref{fig:dzero-ideogram} shows the 1D likelihood as a function of \mtop
along the gray line in the left plot. The result for 0.4 fb$^{-1}$ of data is
$173.7\pm4.4({\rm stat+JES})^{+2.1}_{-2.0}({\rm syst})\GeVcc$.  Dominant
systematic uncertainties are in the modeling of $b$-fragmentation and in the
$b$/light jet energy scale ratio.

\begin{figure}
\begin{center}
\includegraphics[height=2.2in]{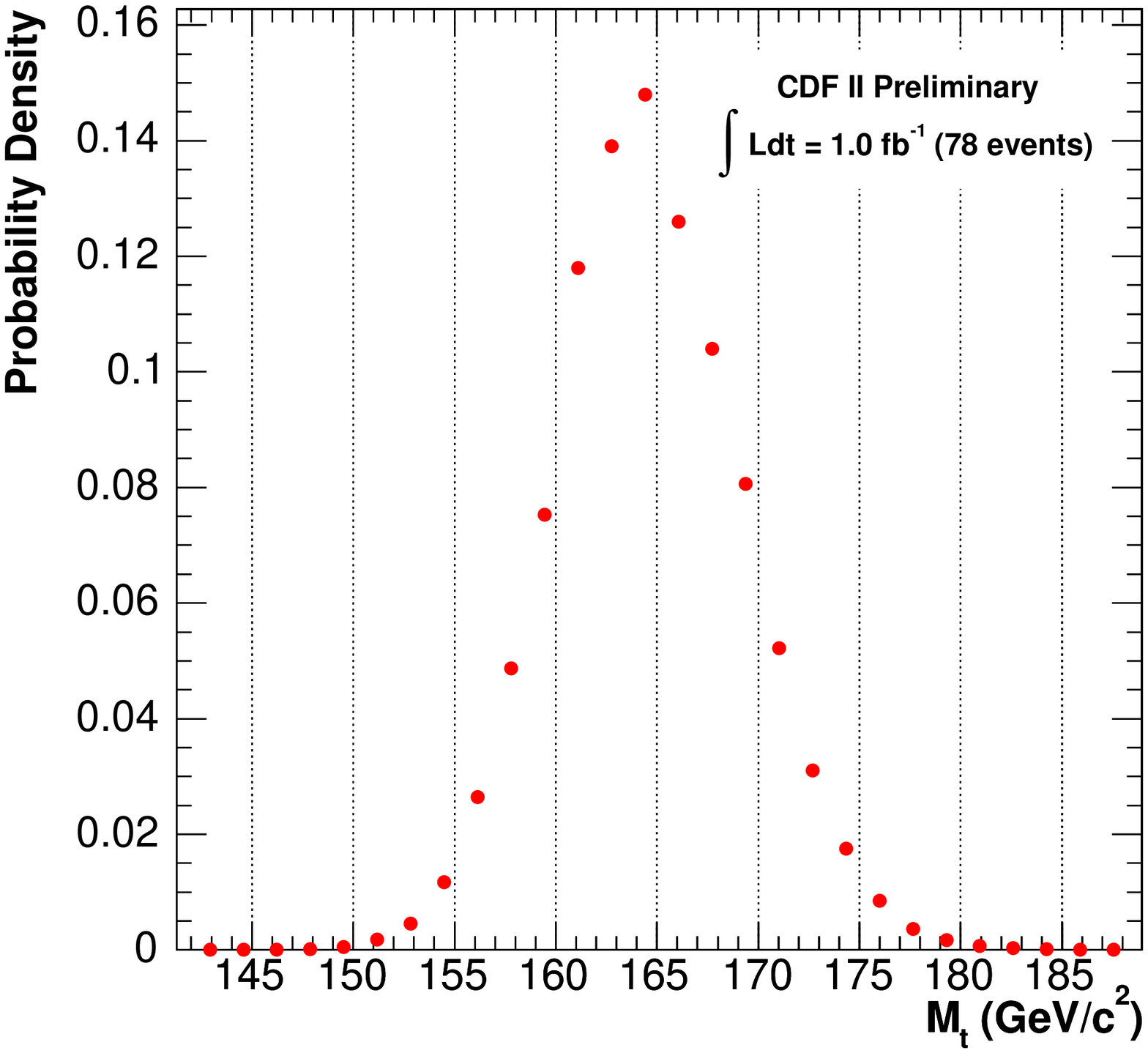}
\includegraphics[height=2.2in]{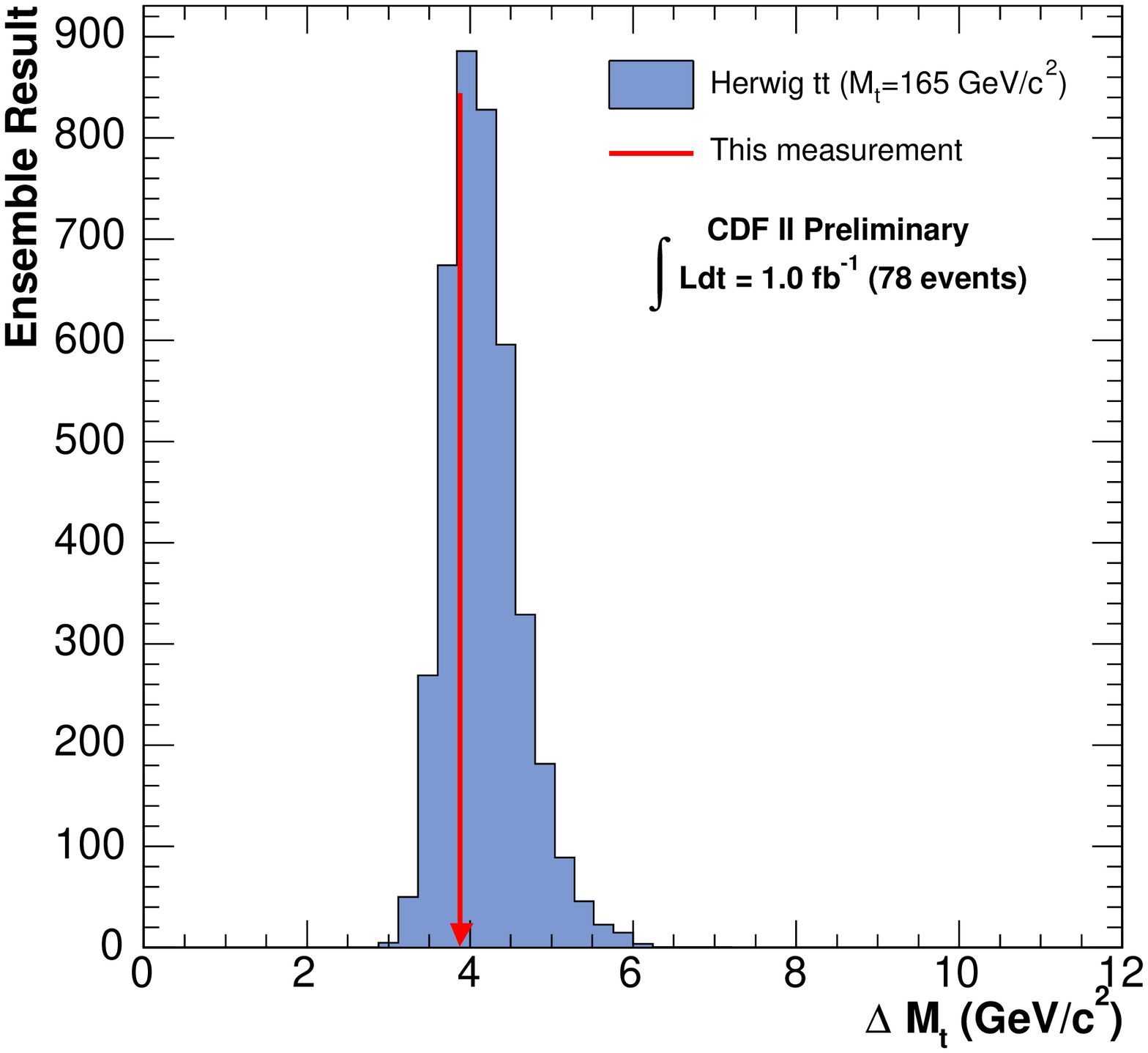}
\caption{Probability and expected error distributions for the CDF dilepton
result.
\label{fig:cdf-dilepton}}
\end{center}
\end{figure}

CDF has applied the ME method to a measurement of the quark top mass in the dilepton
channel \cite{cdfdilepton}.  A plot of the probability as a function of \mtop is
shown on the left in Figure \ref{fig:cdf-dilepton} and the expected error
distribution on the right with the arrow indicating the measurement
uncertainty.  The result for 1 fb$^{-1}$ of data is $164.5\pm3.9({\rm
stat})\pm3.9({\rm syst}) \GeVcc$.  The systematic error is dominated by the
uncertainty in the jet energy scale.

\dzero has measured the top quark mass in the dilepton channel using a template method that
assigns a weight to each neutrino solution based on the agreement between the
calculated transverse momentum of the neutrinos and the observed missing
transverse energy \cite{dzerodilepton}.  The result for 1 fb$^{-1}$ is 
$172.5\pm5.8({\rm stat})\pm5.5({\rm syst})\GeVcc$.  The dominant source of the
systematic error is the jet energy scale uncertainty.

\begin{figure}
\begin{center}
\includegraphics[height=2.2in]{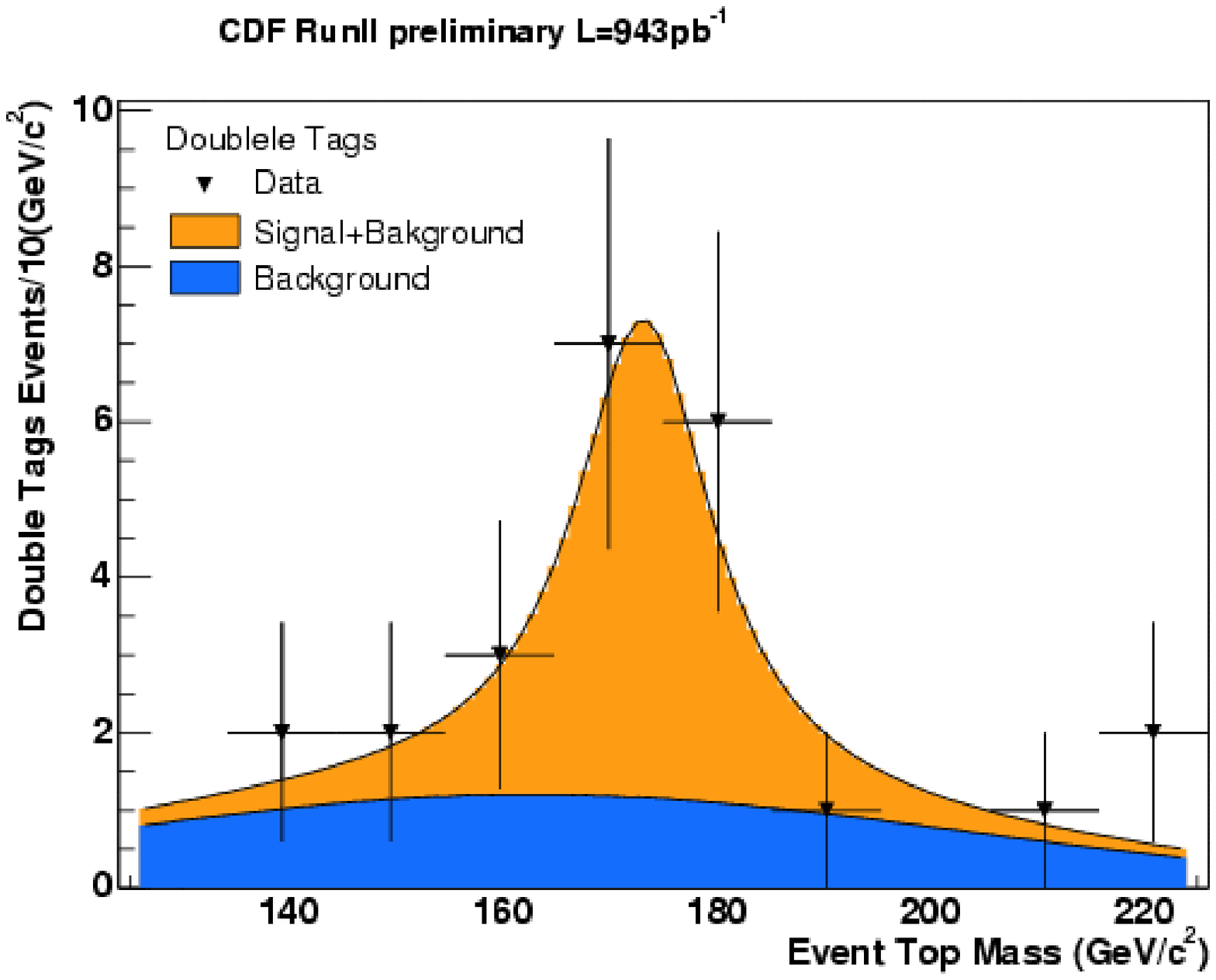}
\includegraphics[height=2.2in]{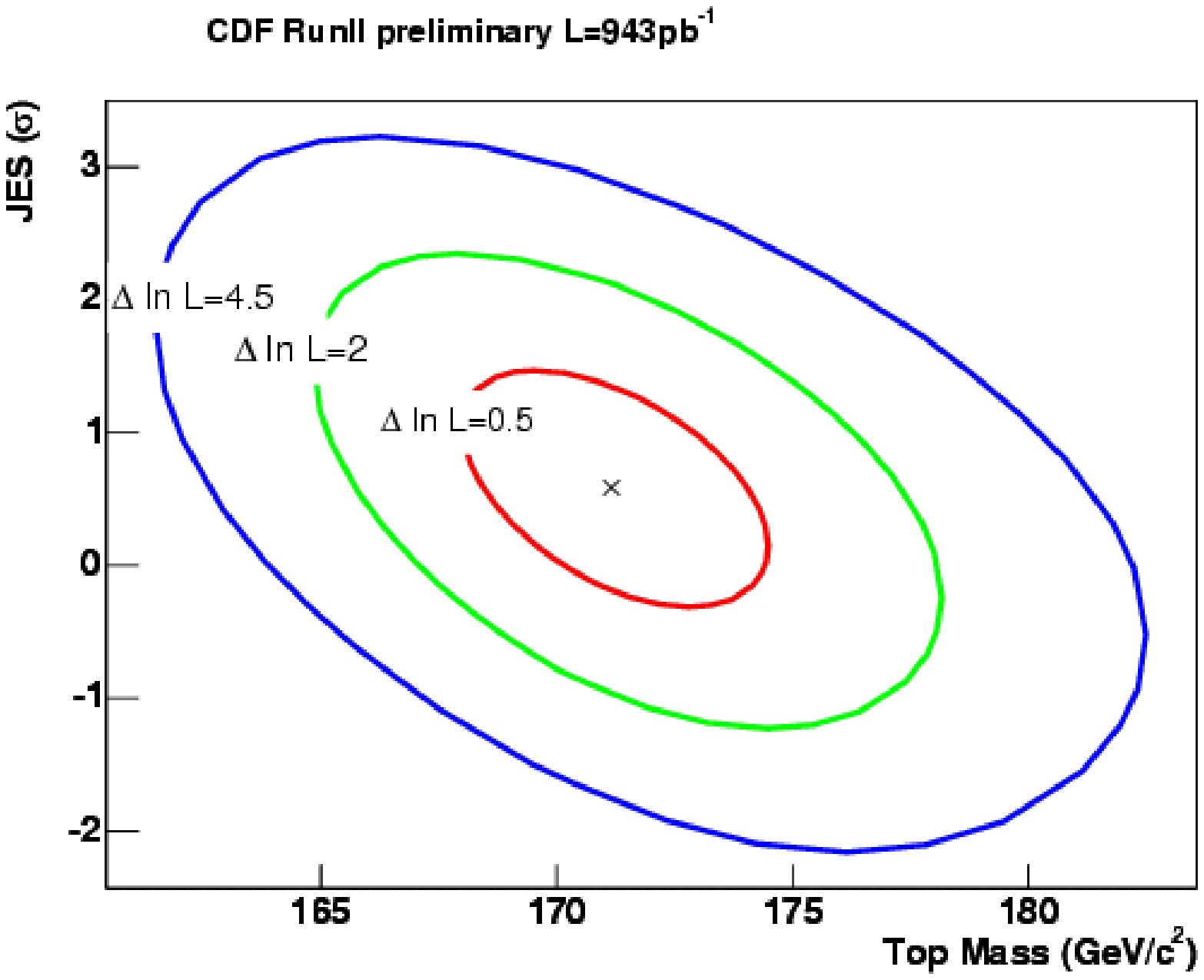}
\caption{Data/MC distributions on the left and mass/$JES$ contours in data
on the right for CDF all jets result
\label{fig:cdf-alljets}}
\end{center}
\end{figure}

CDF has measured the top quark mass in the all jets channel using a combination of template
and ME methods \cite{cdfalljets}.  Instead of using the ME method directly to
measure the top mass, the value determined from the method is used to construct the MC
templates.  Probabilities calculated from the ME are also used in the event
selection process to identify events with high signal probability.  This result
also uses the \mW constraint to perform an in-situ jet energy calibration.  The
left plot in Figure \ref{fig:cdf-alljets} shows a fit of the data distribution
to the MC templates for events with two $b$-tagged jets.  Contours of $JES$ and
\mtop in data are shown on the right in Figure \ref{fig:cdf-alljets}.  The
result for 1 fb$^{-1}$ is $171.1\pm3.7({\rm stat+JES})\pm2.1({\rm
syst})\GeVcc$.  The largest systematic uncertainties are in the simulation of
fragmentation and showering and of final state radiation.

\section{New World Average and Standard Model Fits}

\begin{figure}
\begin{center}
\includegraphics[height=3.7in]{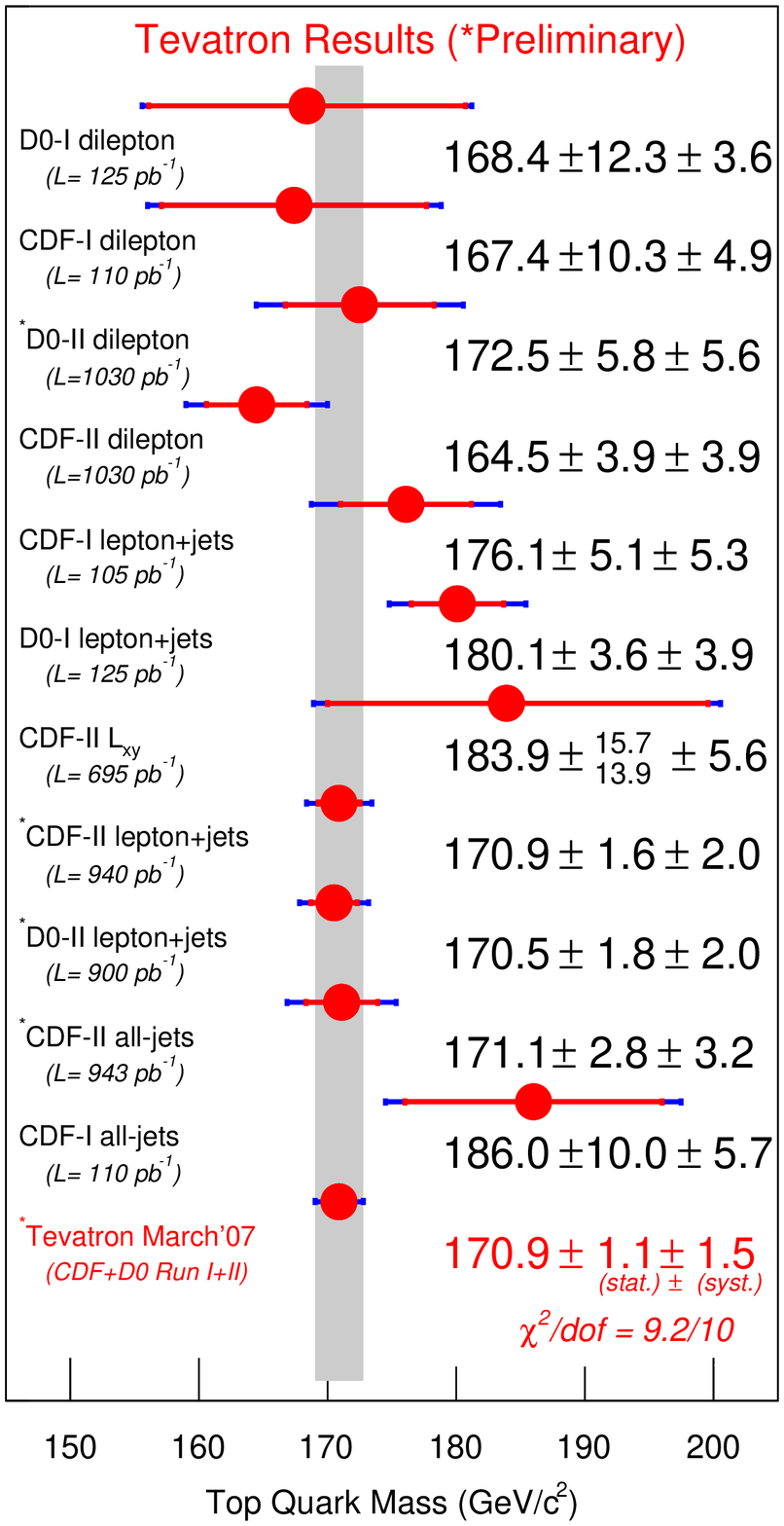}
\includegraphics[height=3.7in]{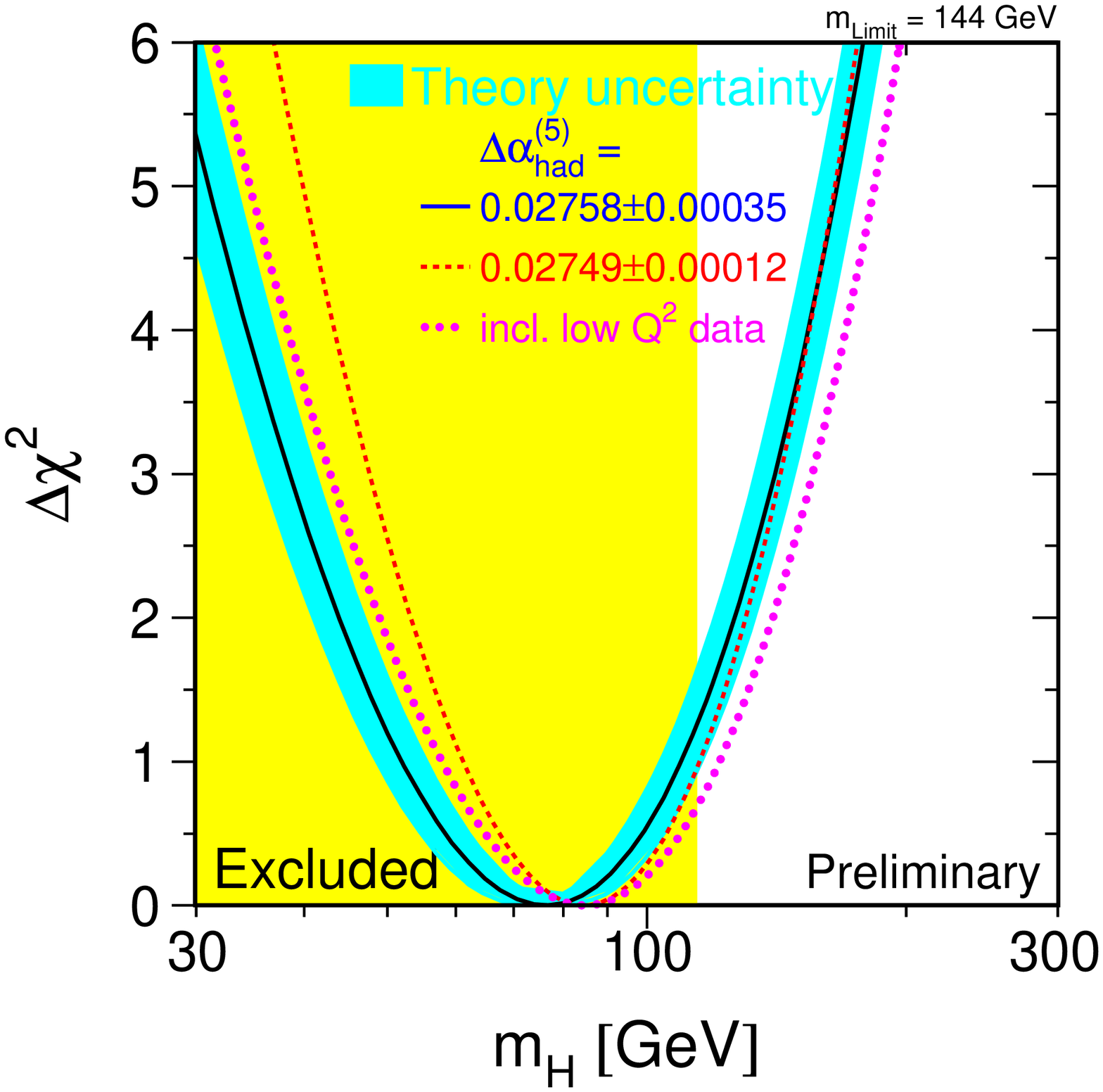}
\caption{New world average \mtop and Standard Model fits.
\label{fig:combination_smfit}}
\end{center}
\end{figure}

From above, the best result of each experiment in each channel is combined with
previous results yielding a new preliminary world average \cite{combination} of
$\mtop=170.9\pm1.1({\rm stat})\pm1.5({\rm syst})\GeVcc$ shown on the left in
Figure \ref{fig:combination_smfit}.  The ME \ljets results from \dzero and CDF 
carry the largest weights in this average of 40\% and 39\%, respectively. This
value is 0.5 \GeVcc lower than the previous world average.  With this new
preliminary result, the top quark mass is now known to a total uncertainty of
1.8 \GeVcc corresponding to a relative precision of 1.1\%.  

This new top quark mass is also combined with other precision electroweak
results in Standard Model fits performed by the LEP Electroweak Working Group
\cite{smfit}. 
The right plot in Figure \ref{fig:combination_smfit} shows the $\Delta\chi^2$
curve resulting from these fits giving  $\mhiggs=76^{+33}_{-24}$ \GeVcc at the
minimum and a 95\% confidence level upper limit of 144 \GeVcc which increases
to 182 \GeVcc when the LEP-2 direct search limit of 114 \GeVcc indicated by the
yellow band is included.

\section{Summary and Conclusions}

A precise determination of the top quark mass is crucial for constraining the
mass of the Higgs boson.  Despite the great challenges involved, precise
measurements are possible through the use of sophisticated measurement
techniques. This talk presented new results based on up to 1 fb$^{-1}$ of data
collected by CDF and \dzero.  Although these results are still dominated by the
\ljets channel, the other two show promise and we hope to see more competitive
results from them in the future.  Combining the new results with previous ones
has yielded a new preliminary world average top quark mass with a total
uncertainty of 1.8 \GeVcc and imposed new constraints on \mhiggs.  As more data
become available at the Tevatron, we can expect statistical uncertainties $<1$
\GeVcc by the end of the Tevatron run at which point the  total uncertainties
will become dominated by the systematic uncertainties.

\end{document}